\documentclass[12pt,preprint]{aastex}

\slugcomment{This manuscript is accepted by \textbf{RAA}}

\shorttitle{Non-similar collapse} \shortauthors{Nejad-Asghar}

\begin{document}

\title{Non-similar collapse of singular isothermal spherical molecular cloud cores with nonzero initial velocities}

\author{Mohsen Nejad-Asghar\altaffilmark{1,2,3}}

\affil{$^1$Department of Physics, University of Mazandaran,
Babolsar, Iran}

\affil{$^2$Research Institute for Astronomy and Astrophysics of
Maragha, Maragha, Iran}

\affil{$^3$Department of Physics, Damghan University, Damghan,
Iran}

\email{nejadasghar@umz.ac.ir}

\begin{abstract}
Theoretically, stars have been formed from the collapse of cores
in the molecular clouds. Historically, the core had been assumed
as an singular isothermal sphere (SIS), and the collapse had been
investigated by a self-similar manner. This is while the rotation
and magnetic field lead to non-symmetric collapse so that a
spheroid shape may be occurred. Here, the resultant of the
centrifugal force and magnetic field gradient is assumed to be in
the normal direction of the rotational axis, and its components
are supposed to be a fraction $\beta$ of the local gravitational
force. In this research, a collapsing SIS core is considered to
find the importance of the parameter $\beta$ for oblateness of
the mass shells which are above the head of the expansion wave. We
apply the Adomian decomposition method to solve the system of
nonlinear partial differential equations because the collapse
does not occur in a spherical symmetry with self-similar
behavior. In this way, we obtain a semi-analytical relation for
the mass infall rate $\dot{M}$ of the shells at the envelope. Near
the rotational axis, the $\dot{M}$ decreases with increasing of
the non-dimensional radius $\xi$, while a direct relation is
observed between $\dot{M}$ and $\xi$ in the equatorial regions.
Also, the values of $\dot{M}$ in the polar regions are greater
than the equatorial values, and this difference is more often at
smaller values of $\xi$. Overall, the results show that before
reaching the head of expansion wave, the visible shape of the
molecular cloud cores can evolve to oblate spheroids. The ratio
of major to minor axes of oblate cores increases with increasing
the parameter $\beta$, and its value can approach to the
apparently observed elongated shapes of cores in the maps of
molecular clouds such as Taurus and Perseus.
\end{abstract}

\keywords{ISM: clouds -- ISM: evolution -- star: formation --
methods: analytical}

\section{Introduction}

Great deals are now known about dense cores in the molecular clouds
that are the progenitors of protostars (e.g., di~Francesco et
al.~2007, Ward-Thompson et al.~2007). We know that, approximately,
all cores in the maps of molecular clouds seem apparently to be
elongated rather than spherical. For example, we can refer to the
recent work of Curtis and Richer~(2010) for two-dimensional
ellipticity of cores in the Perseus molecular cloud, or to the old
report of Myers et al.~(1991) for apparent elongated shapes of cores
in the Taurus maps. Overall, the observations show that, on average,
the ratios of the major to minor axes of cores vary approximately
between $1.2$ to $2$ with a mean value of $\approx 1.6$.

Determining the exact three-dimensional shape of a core from the
apparent observations of the-plane-of-sky is impossible, instead
statistical techniques have to be applied. Several studies have
analyzed the observations in the context of a random distribution
of inclinations to infer that cores are more nearly prolate than
oblate (e.g., Curry~2002, Jones and Basu~2002). This prolate
elongation of cores may be inferred as a remnant of their origin
in filaments (e.g., Hartmann~2002). Also, models with cores
forming from turbulent flows predict random triaxial shapes with
a slight preference for prolateness (Gammie et al.~2003, Li et
al.~2004). Although, some works indicate a preference for prolate
cores, but there are many studies that consistently favour oblate
shapes (e.g., Jones et al.~2001, Goodwin et al.~2002,
Tassis~2007, Offner and Krumholz~2009, Tassis et al.~2009). If
strong magnetic fields are present then collapsing cores are
expected to be oblate (e.g., Galli and Shu~1993, Basu and
Ciolek~2004, Ciolek and Basu~2006), which could also be caused by
strong rotational motion (e.g., Cassen and Moosman~1981, Terebey
et al.~1984).

Historically, in primary theoretical models for the collapse of
cores and formation of star, an isothermal equation of state had
been used, which as a consequence of subsonic communications in
different parts of the cloud, an inverse square profile for
density was appeared (Bodenheimer and Sweigart 1968).
Larson~(1969) and Penston~(1969) were the first to analyze this
inverse square behavior of density profile using the similarity
method. In this case, which had been extended afterwards by
Hunter~(1977), one begin with a static cloud of constant density
and follow the formation of the $r^{-2}$ density profile. In the
opposite case, Shu~(1977) assumed that the density is initially
in inverse square profile SIS core, and constructed the expansion
wave collapse solution to suggest the inside-out collapse
scenario. These two limiting solutions of Hunter and Shu may be
described as fast and slow collapse, respectively, and the
reality may be somewhat in between (McKee and Ostriker~2007).
Since then, a lot of asymptotic solutions and global numerical
simulations have been found and developed in which authors have
approximately considered the effects of three important
mechanisms: turbulence, rotation and magnetic field (see, e.g.,
Hartmann~2009).

In this research, we return to the basic spherical collapse
problem for polytropic spheres, which was used as idealized SIS
similarity solutions by Shu~(1977). In addition, we consider an
initial inward flows, i.e., the conditions observed in some
molecular cloud cores and used by Fatuzzo et al.~(2004) to
investigate its effects on the collapse of SIS. The goal of this
paper is to reexamine the gravitational collapse of SIS with
focus on the oblateness of a core via the effect of rotation and
magnetic field. We suggest that the centrifugal force and
magnetic field pressure lead to oblateness of the envelope of a
molecular core before reaching the collapsing expansion wave.
Since the collapse is not in a spherical symmetric manner, the
similarity method can not be used. Instead, we use the Adomian
decomposition method (Adomain~1994), to solve semi-analytically
the system of differential equations. For this purpose, the
collapse of SIS using the Adomian method is given in section~2.
The non-similar collapse of SIS is investigated in section~3 in
which the oblateness of the core is also obtained. Finally, the
section~4 devotes to summary and conclusion.

\section{Collapse of SIS by Adomian method}

The initial density of a SIS is assumed to be in the form of
inverse square, and the collapse is assumed to be only in the
radial direction. The mass continuity equation is
\begin{equation}\label{mass1}
  \frac{\partial\rho}{\partial t} + \frac{1}{r^2} \frac{\partial}{\partial r} (r^2 \rho u) = 0,
\end{equation}
where $u$ is the radial velocity which follows the force equation,
\begin{equation}\label{mom1}
  \frac{\partial u}{\partial t} + u \frac{\partial u}{\partial r} =
  - \frac{a_s^2}{\rho} \frac{\partial \rho}{\partial r} -g,
\end{equation}
where $a_s$ is the sound speed, and the gravitational
acceleration $g$ follows the Poisson's equation,
\begin{equation}\label{pois1}
  \frac{1}{r^2} \frac{\partial}{\partial r} (r^2 g) = 4 \pi G \rho.
\end{equation}
Choosing the sound speed as the unit of velocity, and
$[t]=1/\sqrt{4 \pi G \rho_c}$ as the unit of time where $\rho_c$
is the density at the center of core, the basic equations
(\ref{mass1})-(\ref{pois1}) can be rewritten as follows
\begin{equation}\label{mass2}
   \frac{\partial\rho}{\partial t} + u \frac{\partial \rho}{\partial
   \xi} + \rho \frac{\partial u}{\partial \xi} + \frac{2}{\xi}
   \rho u = 0,
\end{equation}
\begin{equation}\label{mom2}
   \frac{\partial u}{\partial t} + u \frac{\partial u}{\partial \xi}
  + \frac{1}{\rho} \frac{\partial \rho}{\partial \xi} + g = 0,
\end{equation}
\begin{equation}\label{pois2}
  \frac{1}{\xi^2} \frac{\partial}{\partial \xi} (\xi^2 g) = \rho,
\end{equation}
where $\xi \equiv r/(a_s [t])$ is the nondimensional radius and
the density and gravitational acceleration are
non-dimensionalized by $\rho_c$ and $a_s/[t]$, respectively.

If the cloud is in hydrostatic equilibrium ($u=0$), the equations
(\ref{mass2})-(\ref{pois2}) lead to the Lane-Emden equation which
is well-known in the theory of stellar structure (e.g.,
Chandrasekhar~1939). The Lane-Emden equation can be solved by the
Adomian decomposition method which is found in the appendix~A. The
SIS is a special case of the Lane-Emden equation with inverse
square density which do not conform the boundary conditions.
Here, we choose the initial density as
$\rho(\xi,0)=\Lambda/\xi^2$ where $\Lambda$ is the overdensity
parameter with $\Lambda=2$ for hydrostatic equilibrium. The
boundary condition for gravitational acceleration is $g(0,t)=0$,
and we assume that the initial velocity is inward so that
$u(\xi,0)=-u_\infty$ where $u_\infty$ is constant.

As mentioned in the appendix~A, for using the Adomian
decomposition method, equations (\ref{mass2})-(\ref{pois2}) must
be rewritten as
\begin{equation}\label{mass3}
  L_t \rho  + N_1(\rho,u)=0,
\end{equation}
\begin{equation}\label{mom3}
  L_t u +g + N_2(\rho,u)=0,
\end{equation}
\begin{equation}\label{pois3}
  L_{\xi\xi} g -\rho =0,
\end{equation}
where $L_{t}(\circ) \equiv \frac{d(\circ)}{dt}$ and
$L_{\xi\xi}(\circ)  \equiv \frac{1}{\xi^2} \frac{d}{d\xi}\left(
\xi^2 \frac{d(\circ)}{d\xi} \right)$ are operators, and
\begin{equation}\label{nonlinmass3}
  N_1(\rho,u) \equiv u \frac{\partial \rho}{\partial
   \xi} + \rho \frac{\partial u}{\partial \xi} + \frac{2}{\xi}
   \rho u,
\end{equation}
\begin{equation}\label{nonlinmom3}
  N_2(\rho,u) \equiv u \frac{\partial u}{\partial \xi}
  + \frac{1}{\rho} \frac{\partial \rho}{\partial \xi},
\end{equation}
are the nonlinear terms of the differential equations which can
be written by Adomian series $N_j(\rho,u) = \sum_{n=0}^\infty
A_n^{(j)}$ for $j=1,2$ where
\begin{equation}\label{adompoly1}
  A_n^{(j)}=\frac{1}{n!}\lim_{\lambda\rightarrow 0}\left[ \frac{\partial^n}{\partial \lambda^n}
  N_j\left( \sum_{i=0}^n \rho_i \lambda^i ,  \sum_{i=0}^n u_i \lambda^i \right)
  \right],
\end{equation}
are the Adomian polynomials (Adomian~1994). In this way, the
final solutions are given by the series $\rho=\sum_{n=0}^\infty
\rho_n$, $u=\sum_{n=0}^\infty u_n$ and $g=\sum_{n=0}^\infty g_n$
which the terms can be obtained by the recurrence relations
\begin{equation}\label{recurrho1}
  \rho_{n+1}=-L_t^{-1} A_n^{(1)},
\end{equation}
\begin{equation}\label{recuru1}
  u_{n+1}=-L_t^{-1}(g_n+A_n^{(2)}),
\end{equation}
\begin{equation}\label{recurg1}
  g_{n+1}=L_{\xi\xi}^{-1} \rho_n,
\end{equation}
where $L_t^{-1} (\circ) \equiv \int (\circ) dt$ and
$L_{\xi\xi}^{-1}(\circ) \equiv \int \xi^{-2} \left( \int \xi^2
(\circ) d \xi \right) d \xi$ are the integration operators and
the $n=0$ terms are from initial and boundary conditions: $\rho_0
\equiv \rho(\xi,0)$, $u_0 \equiv u(\xi,0)$ and $g_0 \equiv
g(0,t)$.

Using the mathematical softwares such as Maple (see, appendix~B),
the density and inward velocity are obtained as follows
\begin{equation}\label{rho1}
  \rho= \frac{\Lambda}{\xi^2} \left[ 1- \frac{(\Lambda-2)}{2} \frac{t^2}{\xi^2}
  + u_\infty \frac{2(\Lambda-2)}{3} \frac{t^3}{\xi^3} +
  \frac{(\Lambda-2)(6-2\Lambda+3u_\infty^2)}{4} \frac{t^4}{\xi^4} + O\left(\frac{t^5}{\xi^5}\right)\right]
\end{equation}
\begin{equation}\label{vel1}
  u=-u_\infty-(\Lambda-2)\frac{t}{\xi}\left[ 1- \frac{u_\infty}{2}\frac{t}{\xi} +
  \frac{(6-\Lambda+2u_\infty^2)}{6} \frac{t^2}{\xi^2} - \frac{(24-5\Lambda+3 u_\infty^2)u_\infty}{12}
  \frac{t^3}{\xi^3} \right]+ O\left(\frac{t^5}{\xi^5}\right),
\end{equation}
respectively. In the case of $u_\infty=0$, these results reduce to
the equation (19) of the well-known paper of Shu~(1977) in which
the similarity variable $x$ is replaced by $t/\xi$ (note that
equations (45) and (46) of Fatuzzo et al.~2004 are mistyped).
According to the convergence problem in the series of Adomian
decomposition method (as mentioned in appendix~A), the results
(\ref{rho1}) and (\ref{vel1}) are reliable in the range of $t/\xi
< 1$. Thus, the solutions (\ref{rho1}) and (\ref{vel1}) are
acceptable only for the outer regions from the head of the
expansion wave (i.e., $\xi=t$), and the mass infall rate $\dot{M}
\equiv 4 \pi \xi^2 \rho \mid u \mid$ of the shells in the envelope
can be determined. For the inner regions of expansion wave, the
terms of series (\ref{rho1}) and (\ref{vel1}) are not convergent,
thus, we must use the suitable methods such as piecewise-adaptive
decomposition method (e.g., Ramos~2009) which is beyond the scope
of this research.

\section{Non-similar collapse of SIS}

The cores of molecular clouds rotate and the magnetic fields are
also the non-eliminating parts of this medium. In this section, we
investigate the effect of these mechanisms on the dynamics of
collapsing SIS core. In this case, the infall velocity of matter
in the spherical coordinate have two components as $\mathbf{v}=
\hat{r} u + \hat{\theta} w$ so that the mass continuity equation
is expressed as
\begin{equation}\label{mass4}
   \frac{\partial\rho}{\partial t} + \frac{1}{\xi^2} \frac{\partial}{\partial
   \xi} (\xi^2 \rho u) + \frac{1}{\xi \sin \theta} \frac{\partial}{\partial \theta}
   (\sin \theta \rho w) = 0,
\end{equation}
which is non-dimensionalized according to the units given in
section~2. Here, the effect of the centrifugal force and pressure
of magnetic field, in the envelope of the core, are assumed to be
in the normal direction of the rotational axis as follows
\begin{equation}\label{force4}
\mathbf{F}_{mag+rot} = \hat{r} F_\xi \sin^2 \theta + \hat{\theta}
F_\theta \sin 2 \theta,
\end{equation}
where for simplicity, the components are assumed to be a fraction
of the local gravitational force as $F_\xi= -\beta \partial \phi /
\partial \xi$ and
$F_\theta= -\beta \partial \phi /
\partial \theta$ where the magnetic-rotational parameter $0 \leq \beta \leq 1$
indicates the importance of rotation and magnetic field, and
$\phi$ is the gravitational potential which follows the Poisson's
equation,
\begin{equation}\label{pois4}
  \frac{1}{\xi^2} \frac{\partial}{\partial \xi} (\xi^2
  \frac{\partial \phi}{\partial \xi}) + \frac{1}{\xi^2 \sin \theta}
  \frac{\partial}{\partial \theta} (\sin \theta
  \frac{\partial \phi}{\partial \theta}) = \rho.
\end{equation}
In this way, the force equation have two components as follows
\begin{equation}\label{mom41}
   \frac{\partial u}{\partial t} + u \frac{\partial u}{\partial \xi}
   + \frac{w}{\xi} \frac{\partial u}{\partial \theta} =
  - \frac{1}{\rho} \frac{\partial \rho}{\partial \xi} - (1- \beta \sin^2 \theta)
   \frac{\partial \phi}{\partial \xi},
\end{equation}
\begin{equation}\label{mom42}
  \frac{\partial w}{\partial t} + u \frac{\partial w}{\partial \xi}
   + \frac{w}{\xi} \frac{\partial w}{\partial \theta} =
  - \frac{1}{\rho\xi} \frac{\partial \rho}{\partial \theta} - (1- \beta \sin 2 \theta)
   \frac{1}{\xi} \frac{\partial \phi}{\partial \theta}.
\end{equation}

Since the components of gravitational force, $\partial \phi /
\partial \xi$ and $\partial \phi / \partial \theta$, are zero at
the center of the core, the boundary condition of the
gravitational potential at $\xi=0$ is assumed to be
$\phi(0,t)=0$. At the beginning of collapse, the components of
velocity are assumed to be $u(r,0)=-u_\infty$ and $w(r,0)=0$, and
the initial density is assumed as the density of SIS:
$\rho(\xi,0)= \Lambda/ \xi^2$. We rewrite the the equations
(\ref{mass4}), (\ref{pois4}), (\ref{mom41}) and (\ref{mom42}) in
the Adomian form as follows
\begin{equation}\label{adommass4}
  L_t \rho  + N_1(\rho,u,w)=0,
\end{equation}
\begin{equation}\label{adompois4}
  L_{\xi\xi} \phi + \frac{1}{\xi^2} \frac{\partial^2 \phi}{\partial \theta^2}
  + \frac{\cot \theta}{\xi^2} \frac{\partial \phi}{\partial \theta} -\rho =0,
\end{equation}
\begin{equation}\label{adommom41}
  L_t u + (1- \beta \sin^2 \theta)
   \frac{\partial \phi}{\partial \xi} + N_2(\rho,u,w)=0,
\end{equation}
\begin{equation}\label{adommom42}
  L_t w + (1- \beta \sin 2 \theta)
   \frac{1}{\xi} \frac{\partial \phi}{\partial \theta} + N_3(\rho,u,w)=0,
\end{equation}
respectively, where the nonlinear terms are
\begin{equation}\label{nonlinmass4}
  N_1(\rho,u,w) \equiv u \frac{\partial \rho}{\partial
   \xi} + \rho \frac{\partial u}{\partial \xi} + \frac{2}{\xi}
   \rho u + \frac{w}{\xi} \frac{\partial \rho}{\partial
   \theta} + \frac{\rho}{\xi} \frac{\partial w}{\partial \theta} + \frac{\cot \theta}{\xi}
   \rho w,
\end{equation}
\begin{equation}\label{nonlinmom41}
  N_2(\rho,u,w) \equiv u \frac{\partial u}{\partial \xi}
  + \frac{w}{\xi} \frac{\partial u}{\partial \theta}
  + \frac{1}{\rho} \frac{\partial \rho}{\partial \xi},
\end{equation}
\begin{equation}\label{nonlinmom42}
  N_3(\rho,u,w) \equiv u \frac{\partial w}{\partial \xi}
  + \frac{w}{\xi} \frac{\partial w}{\partial \theta}
  + \frac{1}{\rho\xi} \frac{\partial \rho}{\partial \theta}.
\end{equation}
In this way, by appointment the Adomian polynomials,
\begin{equation}\label{adompoly2}
  A_n^{(j)}=\frac{1}{n!}\lim_{\lambda\rightarrow 0} \left[ \frac{\partial^n}{\partial \lambda^n}
  N_j\left( \sum_{i=0}^n \rho_i \lambda^i ,  \sum_{i=0}^n u_i \lambda^i,
  \sum_{i=0}^n w_i \lambda^i \right)
  \right],
\end{equation}
for $j=1,2,3$, we receive to the recurrence relations as follows
\begin{equation}\label{recurrho2}
  \rho_{n+1}=-L_t^{-1} A_n^{(1)},
\end{equation}
\begin{equation}\label{recurphi2}
  \phi_{n+1}=-L_{\xi\xi}^{-1} \left[ \frac{1}{\xi^2} \frac{\partial^2 \phi_n}{\partial \theta^2}
  + \frac{\cot \theta}{\xi^2} \frac{\partial \phi_n}{\partial \theta}-\rho_n \right],
\end{equation}
\begin{equation}\label{recuru2}
  u_{n+1}=-L_t^{-1} \left[(1- \beta \sin^2 \theta)
   \frac{\partial \phi_n}{\partial \xi}+A_n^{(2)} \right],
\end{equation}
\begin{equation}\label{recurw2}
  w_{n+1}=-L_t^{-1}\left[(1- \beta \sin 2 \theta)
   \frac{1}{\xi} \frac{\partial \phi_n}{\partial
   \theta}+A_n^{(3)}\right],
\end{equation}
where the $n=0$ terms are given by the initial and boundary
conditions. Thus, we can obtain the terms of the series
$\rho=\sum_{n=0}^\infty \rho_n$, $u=\sum_{n=0}^\infty u_n$,
$w=\sum_{n=0}^\infty w_n$ and $\phi=\sum_{n=0}^\infty \phi_n$
with a mathematical software such as Maple (see, appendix~B).
Here, we turn our attention to the mass infall rate, $\dot{M}
\equiv 4 \pi \xi^2 \rho \mid u \mid$, at the envelope of the core
($\xi > t$) where the Adomian decomposition method is reliable.
The result is
\begin{equation}\label{mdot}
  \dot{M}= 4 \pi \Lambda u_\infty + 4 \pi \Lambda (\Lambda -2 -
  \Lambda \beta \sin^2 \theta) \frac{t}{\xi} - 4 \pi \Lambda
   (\Lambda -2 - \Lambda \beta \sin^2 \theta) u_\infty
   \frac{t^2}{\xi^2} + O\left( \frac{t^3}{\xi^3} \right),
\end{equation}
which its values for $\Lambda = 2.1$, $u_\infty=0.1$ and
$\beta=0.1$, at time $t=1/6$ are shown in Fig.\ref{mdotab}.

\begin{figure} \epsscale{.6} \plotone{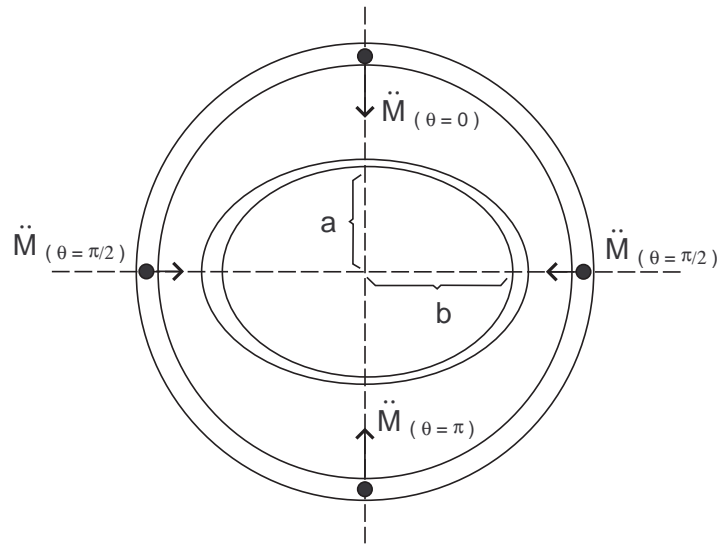}
\caption{Schematic diagram for mass infall rate which leads to
the formation of a spheroidal shell from an initial spherical
shell. \label{scheme}}
\end{figure}

\begin{figure}
\epsscale{.35} \center \plotone{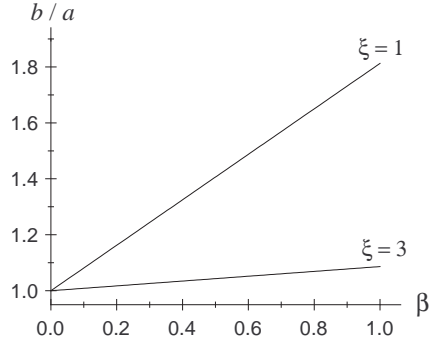}\\{(a)}\\
\epsscale{.35} \center \plotone{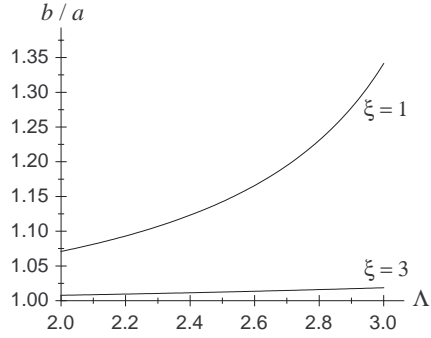}\\{(b)}\\
\epsscale{.35} \center \plotone{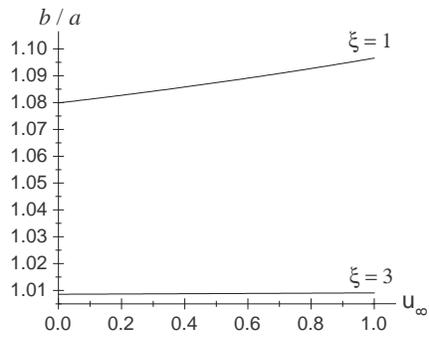}\\{(c)}\\
\caption{The ratio of major to minor axes of spheroidal shells of
a core, at time $t=1/6$, versus (a) magnetic-rotational parameter
$\beta$ with $\Lambda=2.1$ and $u_\infty=0.1$, (b) overdensity
parameter $\Lambda$ with $\beta=0.1$ and $u_\infty=0.1$, and (c)
inward initial velocity $u_\infty$ with $\Lambda=2.1$ and
$\beta=0.1$. \label{btoa}}
\end{figure}

Since the infall rates at the polar and equatorial regions are
different, an initial spherical shell at the outer region of
expansion wave, will be spheroid by time, as schematically shown
in Fig.~\ref{scheme}. In the first order approximation of mass
infall rate (i.e., in the order of $t/\xi$), ratio of major to
minor axes of the spheroid can be approximated as
\begin{equation}\label{btoa1}
  \frac{b}{a} \equiv \frac{\xi- u_\infty t - \frac{1}{2} \ddot{M}_{(\theta=\pi/2)}
  t^2}{\xi-u_\infty t - \frac{1}{2} \ddot{M}_{(\theta=0)}
  t^2}.
\end{equation}
Inserting $\ddot{M} \simeq 4 \pi \Lambda (\Lambda -2 - \Lambda \beta
\sin^2 \theta)/\xi$, which is obtained from equation (\ref{mdot}),
we have
\begin{equation}\label{btoa2}
   \frac{b}{a}= \frac{1-u_\infty \frac{t}{\xi} - 2\pi\Lambda (\Lambda -2 - \Lambda \beta)
   \frac{t^2}{\xi^2}}{1-u_\infty \frac{t}{\xi} - 2\pi\Lambda (\Lambda -2)
   \frac{t^2}{\xi^2}},
\end{equation}
which is depicted in Fig.~\ref{btoa} at time $t=1/6$.

\section{Summary and conclusions}

Stars have been formed from collapse of cores in the molecular
clouds. A basic standard model for collapse of cores assumes a
SIS, in which the collapse occurs inside-out accompanied with an
expansion wave. We know that not only the molecular cores rotate,
but also the magnetic fields affect on their dynamics. If we
assume that the effect of centrifugal force and magnetic pressure
are in the normal direction of the rotational axis, we expect
that the shape of the envelope of core (outer regions from the
head of expansion wave) be modified to a spheroid. In this case
which the collapse is in a non-symmetric manner, we used the
Adomian decomposition method to solve the differential equations.
In the appendix~A, we solved the well-known Lane-Emden equation to
find that the Adomian method is applicable. Then, in section~2,
the collapse of a SIS core is investigated. The results show that
the Adomian decomposition method presents convergent solutions
only for the regions outer than the head of the expansion wave
(i.e., envelope).

\begin{figure}
\epsscale{.55} \center \plotone{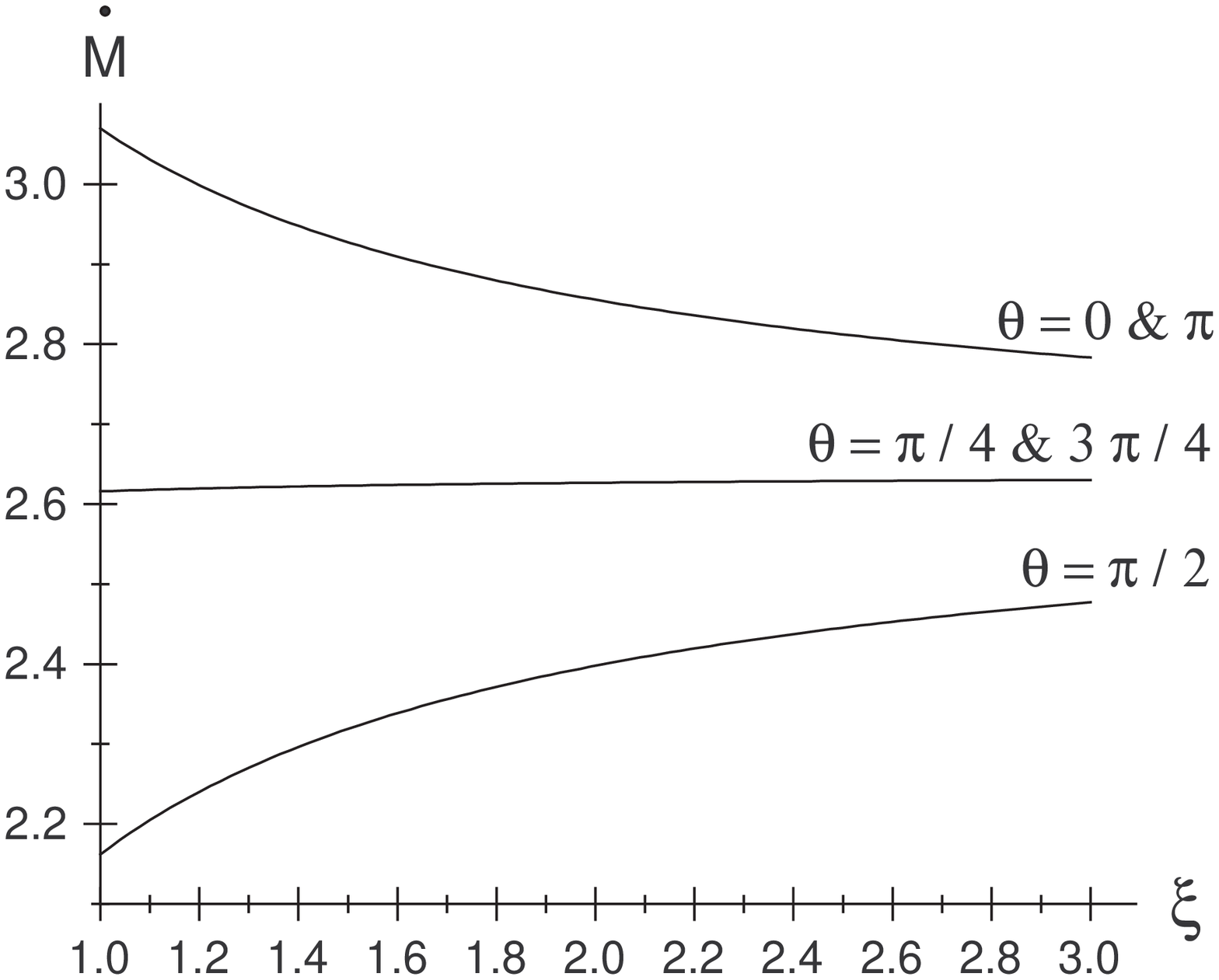}\\{(a)}\\
\epsscale{.55} \center \plotone{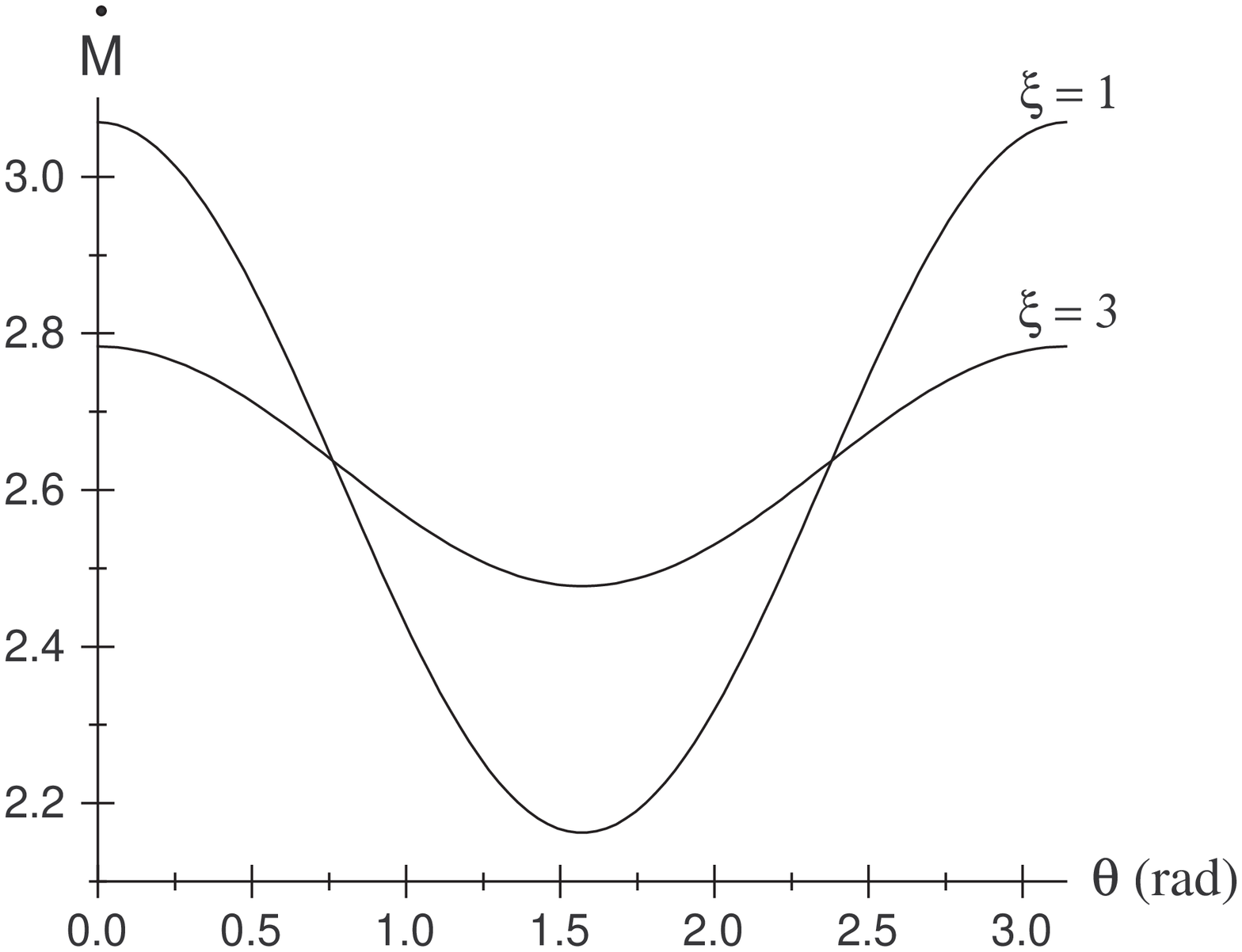}\\{(b)}\\
\caption{Mass infall rate versus (a) nondimensional length $\xi$,
and (b) polar angle $\theta$, with $\Lambda = 2.1$, $u_\infty=0.1$
and $\beta=0.1$, at time $t=1/6$. \label{mdotab}}
\end{figure}

In section~3, non-similar collapse of a SIS core, which is
affected by the centrifugal force and magnetic pressure, is solved
by the Adomian decomposition method. The mass infall rate of the
shells in the envelope of the core is obtained and depicted in
Fig.\ref{mdotab} with typical values of the parameters. According
to Fig.~\ref{mdotab}(a), the mass infall rate in the polar regions
(near the rotational axis) decreases with increasing of radius,
while in equatorial regions, the mass infall rate in the regions
far from the head of expansion wave is greater than regions near
to it. Fig.~\ref{mdotab}(b) shows that the mass infall rate in
equatorial regions are less than polar regions, thus, before
reaching the expansion wave, shape of the envelope converts to a
spheroid as shown schematically in the Fig.~\ref{scheme}.

In this way, according to mass infall rate, we determined the
ratio of major to minor axes of the spheroid, and its values are
shown in the Fig.~\ref{btoa} with typical values of the
parameters. We see that the ratio of major to minor axes of
spheroid do not strongly depend on the initial inward velocity
$u_\infty$, while its value depends on the overdensity parameter
$\Lambda$ and magnetic-rotational parameter $\beta$. Thus, the
results show that the magnetic field and rotation lead to
non-symmetric collapse so that before reaching the head of
expansion wave, the apparent shape of the envelope will be
converted to the triaxial oblate spheroids. Also, the results
show that the ratio of major to minor axes of the spheroids can
reach to the values which are consistent with the observed maps
of the cores in the molecular clouds such as Taurus and Perseus.

\section*{Acknowledgments}
This work has been supported by Research Institute for Astronomy
and Astrophysics of Maragha (RIAAM). Some parts of this article
are presented in the thesis of Zeynab Mirzaii, who is MSc student
in the Damghan University, and the author is her supervisor.

\appendix

\section{Solving the Lane-Emden equation by Adomian decomposition method}

A spherical polytropic gas, in hydrostatic equilibrium, satisfies
the equation
\begin{equation}\label{equilib}
  \frac{1}{\rho} \frac{dp}{dr}= - \frac{GM(r)}{r^2},
\end{equation}
where $M(r)$ is the mass inside the radius $r$ and the polytropic
pressure is $p=\kappa \rho^\Gamma$, which in the isothermal case,
we have $\kappa=a_s^2$ and $\Gamma=1$, where $a_s$ is the
isothermal sound speed. Substituting $\rho=\rho_c \exp (-\psi)$
into equation (\ref{equilib}), we have
\begin{equation}\label{polylanem}
   \frac{1}{\xi^2} \frac{\partial}{\partial \xi} \left[ \xi^2 e^{(1-\Gamma)\psi} \frac{d\psi}{d\xi}
   \right]=e^{-\psi},
\end{equation}
where $\xi \equiv \left( \frac{4\pi G}{\kappa \Gamma
\rho_c^{\Gamma-2}} \right)^{1/2}$ is the nondimensional radius. In
the isothermal case, the equation (\ref{polylanem}) reduces to the
well-known Lane-Emden equation. The boundary conditions are
$\psi_{(\xi=0)}=0$ and $\left(d\psi / d\xi \right)_{\xi=0} =0$.
Thus, the equation (\ref{polylanem}) can straightforwardly be
solved by numerical methods such as fourth-order Runge-Kutta.

Here, we solve the equation (\ref{polylanem}) by the Adomian
decomposition method, then we compare the result with the
numerical method. For this purpose, we rewrite the equation
(\ref{polylanem}) as follows
\begin{equation}\label{adomlanem}
  L_{\xi\xi}\psi+ N(\psi)=0,
\end{equation}
where $L_{\xi\xi}(\circ)  \equiv \frac{1}{\xi^2}
\frac{d}{d\xi}\left( \xi^2 \frac{d(\circ)}{d\xi} \right)$ is the
operator, and nonlinear terms of the differential equation are
assembled in the function $N(\psi) \equiv (1-\Gamma)\left(
\frac{d\psi}{d\xi} \right)^2 - e^{(\Gamma-2) \psi}$. The basis of
the Adomian decomposition method is to replace the function
$\psi$ by a series $\psi= \sum_{n=0}^\infty \psi_n$, and the
nonlinear term $N(\psi)$ by a Taylor expansion series $N(\psi)=
\sum_{n=0}^\infty A_n$ where $A_n$ are Adomian polynomials
\begin{eqnarray}\label{adompol1}
 \nonumber &&A_0=N(\psi_0),\\
 \nonumber &&A_1= \psi_1 \left( \frac{d\psi}{d\xi}
 \right)_{\psi=\psi_0},\\
 &&A_2= \psi_2 \left( \frac{d\psi}{d\xi}
 \right)_{\psi=\psi_0}+ \frac{1}{2!} \psi_1^2 \left( \frac{d^2\psi}{d\xi^2}
 \right)_{\psi=\psi_0},
\end{eqnarray}
and so on (Adomian~1994), which can be expressed in the general
form
\begin{equation}\label{adompol2}
  A_n=\frac{1}{n!}\lim_{\lambda\rightarrow 0} \left[ \frac{d^n}{d \lambda^n}
  N\left( \sum_{i=0}^n \psi_i \lambda^i \right)
  \right],
\end{equation}
where $\lambda$ is only a dummy variable which is inserted to
recover the equations (\ref{adompol1}).

\begin{figure} \epsscale{.8} \plotone{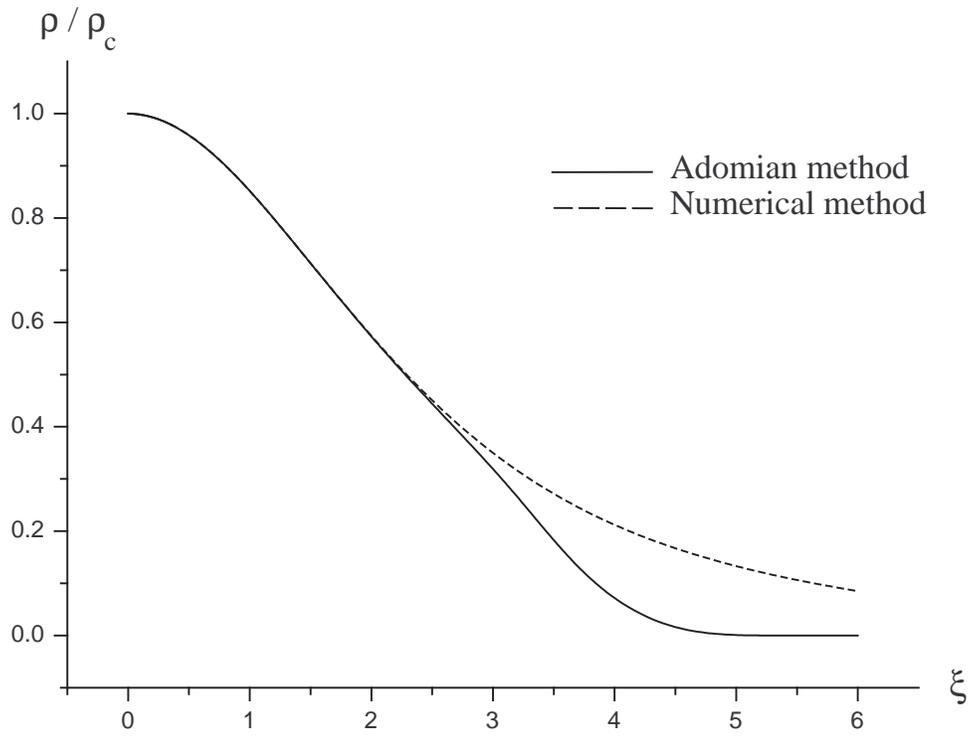}
\caption{Ratio of density to the central density of an isothermal
spherical core in the hydrostatic equilibrium. \label{lanem}}
\end{figure}

In this way, the equation (\ref{polylanem}) leads to a recurrence
relation
\begin{equation}\label{reclanem}
  \psi_{n+1} = -L_{\xi\xi}^{-1} A_n,
\end{equation}
where $L_{\xi\xi}^{-1}(\circ) \equiv \int \xi^{-2} \left( \int
\xi^2 (\circ) d \xi \right) d \xi$ is the integration operator and
$\psi_0 \equiv \psi_{(\xi=0)}$. Thus, we can determine the
$\psi_n$s with help of any mathematical softwares such as Maple
(see, appendix~B). The result is a series as
\begin{equation}\label{sollanem}
  \psi= a_2 \xi^2 + a_4 \xi^4 + a_6 \xi^6 + ...,
\end{equation}
where the coefficient are
\begin{eqnarray}\label{coeflanem}
  \nonumber && a_2=+\frac{1}{3!},\\
  \nonumber && a_4=- \frac{1}{3}(8-5\Gamma) \frac{1}{5!},\\
  && a_6= +\frac{2}{3}\left[ 4(2-\Gamma)^2 +5(2-\Gamma)(1-\Gamma)
  + \frac{8}{3}(1-\Gamma)^2 \right] \frac{1}{7!},
\end{eqnarray}
and so on. In isothermal case ($\Gamma=1$), this result reduces
to the result of Wazwaz~(2001). In Fig.~\ref{lanem}, the ratio of
density to central density, $\frac{\rho}{\rho_c}=e^{-\psi}$, in
the isothermal case, which is obtained from three terms of the
series (\ref{coeflanem}) is compared to the numerical results,
which are obtained from the fourth-order Runge-Kutta method. As
can be seen, the Adomian decomposition method gives reliable
solutions for $\xi<2$. This is because of using the Taylor
expansion which destroys the convergence of the series at large
$\xi$ (Liao~2003, Singh et al.~2009). Substituting the
appropriate values of the typical molecular clouds, we have
\begin{equation}\label{xilanem}
  \xi= \frac{r_{(\mathrm{pc})}}{0.03\mathrm{pc} \left( \frac{T}{10\mathrm{K}}
  \right)^{1/2} \left( \frac{n_c}{10^4 \mathrm{cm^{-3}}}
  \right)^{-1/2}},
\end{equation}
where $n_c$ is the number density at the center of the core.
Thus, in the hydrostatic equilibrium of a typical molecular cloud
core, the results which are obtained by the Adomian decomposition
method, are usable only for radii less than $\sim
0.06\mathrm{pc}$.

\section{Maple programs}

\begin{itemize}
  \item The Maple program for the Lane-Emden equation is as follows:\\
\verb"> psi[0]:= 0:"\\
\verb"> n:= 0:"\\
\verb"> N:= proc (psi) options operator, arrow;"\\
\verb"       (1-Gamma)*(diff(psi,xi))^2-exp((Gamma-2)*psi)"\\
\verb"      end proc:"\\
\verb"> psi[1]:= -int(xi^(-2)*int( xi^2*N(psi[0]),xi),xi):"\\
\verb"> for n from 1 to 7 do"\\
\verb"    N:= proc (psi) options operator, arrow;"\\
\verb"         (1-Gamma)*(diff(psi,xi))^2-exp((Gamma-2)*psi)"\\
\verb"        end proc:"\\
\verb"    psisum:= sum(psi[m]*lambda[n]^m,m=0..n):"\\
\verb"    A[n]:= (diff(N(psisum),$(lambda[n],n)))/factorial(n):"\\
\verb"    lambda[n]:= 0:"\\
\verb"    psi[n+1]:= -int(xi^(-2)*int(xi^2*A[n],xi),xi):"\\
\verb"  end do:"\\
\verb"> psitotal:= collect(sum(psi[m],m=0..n),xi);"\\

  \item The Maple program for the collapse of SIS is as follows:\\
\verb"> g[0]:= 0:"\\
\verb"> rho[0]:= Lambda/xi^2:"\\
\verb"> u[0]:= -u[infinity]:"\\
\verb"> n:= 0:"\\
\verb"> N1:= proc (rho,u) options operator, arrow;"\\
\verb"        u*diff(rho,xi)+rho*diff(u,xi)+2*u*rho/xi"\\
\verb"       end proc:"\\
\verb"> N2:= proc (rho,u) options operator, arrow;"\\
\verb"        u*diff(u,xi)+rho^(-1)*diff(rho,xi)"\\
\verb"       end proc:"\\
\verb"> g[1]:= xi^(-2)*int(xi^2*rho[0],xi):"\\
\verb"> rho[1]:= -int(N1(rho[0],u[0]),t):"\\
\verb"> u[1]:= -int(g[0]+N2(rho[0],u[0]),t):"\\
\verb"> for n from 1 to 7 do"\\
\verb"    N1:= proc (rho,u) options operator, arrow;"\\
\verb"          u*diff(rho,xi)+rho*diff(u,xi)+2*u*rho/xi"\\
\verb"         end proc:"\\
\verb"    N2:= proc (rho,u) options operator, arrow;"\\
\verb"          u*diff(u,xi)+rho^(-1)*diff(rho,xi)"\\
\verb"         end proc:"\\
\verb"    rhosum:= sum(rho[m]*lambda[n]^m,m=0..n):"\\
\verb"    usum:= sum(u[m]*lambda[n]^m,m=0..n):"\\
\verb"    A1[n]:= (diff(N1(rhosum,usum),$(lambda[n],n)))/factorial(n):"\\
\verb"    A2[n]:= (diff(N2(rhosum,usum),$(lambda[n],n)))/factorial(n):"\\
\verb"    lambda[n]:= 0:"\\
\verb"    g[n+1]:= xi^(-2)*int(xi^2*rho[n],xi):"\\
\verb"    rho[n+1]:= -int(A1[n],t):"\\
\verb"    u[n+1]:= -int(g[n]+A2[n],t):"\\
\verb"  end do:"\\
\verb"> rhototal:= collect(sum(rho[m],m=0..n),t);"\\
\verb"> utotal:= collect(sum(u[m],m=0..n),t);"\\
\verb"> gtotal:= collect(sum(g[m],m=0..n),t);"\\

  \item The Maple program for the non-similar collapse of SIS is as follows:\\
\verb"> Phi[0]:= 0:"\\
\verb"> rho[0]:= Lambda/xi^2:"\\
\verb"> u[0]:= -u[infinity]:"\\
\verb"> w[0]:= 0:"\\
\verb"> n:= 0:"\\
\verb"> N1:= proc (rho,u,w) options operator, arrow;"\\
\verb"        u*diff(rho,xi)+rho*diff(u,xi)+2*u*rho/xi+rho/xi*diff(w,theta)"\\
\verb"        +w/xi*diff(rho,theta)+cot(theta)*rho*w/xi"\\
\verb"       end proc:"\\
\verb"> N2:= proc (rho,u,w) options operator, arrow;"\\
\verb"        u*diff(u,xi)+w/xi*diff(u,theta)+rho^(-1)*diff(rho,xi)"\\
\verb"       end proc:"\\
\verb"> N3:= proc (rho,u,w) options operator, arrow;"\\
\verb"        u*diff(w,xi)+w/xi*diff(w,theta)+rho^(-1)*diff(rho,theta)/xi"\\
\verb"       end proc:"\\
\verb"> Phi[1]:= -int(xi^(-2)*int(xi^2*(diff(Phi[0],theta,theta)/xi^2"\\
\verb"            +cot(theta)*diff(Phi[0],theta)/xi^2-rho[0]),xi),xi):"\\
\verb"> rho[1]:= -int(N1(rho[0],u[0],w[0]),t):"\\
\verb"> u[1]:= -int((1-beta*(sin(theta))^2)*diff(Phi[0],xi)"\\
\verb"         +N2(rho[0],u[0],w[0]),t):"\\
\verb"> w[1]:= -int((1-beta*sin(2*theta))*diff(Phi[0],theta)/xi"\\
\verb"         +N3(rho[0],u[0],w[0]),t):"\\
\verb"> for n from 1 to 7 do"\\
\verb"    N1:= proc (rho,u,w) options operator, arrow;"\\
\verb"          u*diff(rho,xi)+rho*diff(u,xi)+2*u*rho/xi+rho/xi*diff(w,theta)"\\
\verb"          +w/xi*diff(rho,theta)+cot(theta)*rho*w/xi"\\
\verb"         end proc:"\\
\verb"    N2:= proc (rho,u,w) options operator, arrow;"\\
\verb"          u*diff(u,xi)+w/xi*diff(u,theta)+rho^(-1)*diff(rho,xi)"\\
\verb"         end proc:"\\
\verb"    N3:= proc (rho,u,w) options operator, arrow;"\\
\verb"          u*diff(w,xi)+w/xi*diff(w,theta)+rho^(-1)*diff(rho,theta)/xi"\\
\verb"         end proc:"\\
\verb"    rhosum:=sum(rho[m]*lambda[n]^m,m=0..n):"\\
\verb"    usum:=sum(u[m]*lambda[n]^m,m=0..n):"\\
\verb"    wsum:=sum(w[m]*lambda[n]^m,m=0..n):"\\
\verb"    A1[n]:= (diff(N1(rhosum,usum,wsum),$(lambda[n],n)))/factorial(n):"\\
\verb"    A2[n]:= (diff(N2(rhosum,usum,wsum),$(lambda[n],n)))/factorial(n):"\\
\verb"    A3[n]:= (diff(N3(rhosum,usum,wsum),$(lambda[n],n)))/factorial(n):"\\
\verb"    A4[n]:= (diff(N4(rhosum,usum,wsum),$(lambda[n],n)))/factorial(n):"\\
\verb"    lambda[n]:=0:"\\
\verb"    Phi[n+1]:=-int(xi^(-2)*int(xi^2*(diff(Phi[n],theta,theta)/xi^2"\\
\verb"               +cot(theta)*diff(Phi[n],theta)/xi^2-rho[n]),xi),xi):"\\
\verb"    rho[n+1]:=-int(A1[n],t):"\\
\verb"    u[n+1]:=-int((1-beta*(sin(theta))^2)*diff(Phi[n],xi)+A2[n],t):"\\
\verb"    w[n+1]:=-int((1-beta*sin(2*theta))*diff(Phi[n],theta)/xi+A3[n],t):"\\
\verb"  end do:"\\
\verb"> rhototal:=collect(sum(rho[m],m=0..n),t):"\\
\verb"> utotal:=collect(sum(u[m],m=0..n),t):"\\
\verb"> wtotal:=collect(sum(w[m],m=0..n),t):"\\
\verb"> phitotal:=collect(sum(Phi[m],m=0..n),t):"\\
\verb"> mdot:=collect(4*Pi*xi^2*(rhototal)*(-utotal),t);"\\
\end{itemize}


\clearpage

\clearpage

\clearpage

\clearpage

\end{document}